
\input phyzzx
\hoffset=0.2truein
\voffset=0.1truein
\hsize=6truein
\def\TITLEPAGE{\frontpagetrue}
\def\CALT#1{\hbox to\hsize{\tenpoint \baselineskip=12pt
        \hfil\vtop{
        \hbox{\strut CALT-68-#1}
        \hbox{\strut DOE RESEARCH AND}
        \hbox{\strut DEVELOPMENT REPORT}}}}

\def\CALTECH{
        \address{California Institute of Technology,
Pasadena, CA 91125}}
\def\TITLE#1{\vskip .5in \centerline{\fourteenpoint #1}}

\def\ABSTRACT#1{\vskip .2in \vfil \centerline{\twelvepoint
\bf Abstract}
        #1 \vfil}
\def\ENDTITLEPAGE{\vfil\eject\pageno=1}
\hfuzz=5pt
\tolerance=10000
\TITLEPAGE
\CALT{1753}
\TITLE{On Detecting Discrete Cheshire
Charge\footnote\dagger{This work supported in part
by the US Department of Energy under Contract No. DE-AC03-
81-ER40050}}
\vskip 15pt
\centerline{Martin Bucher, Kai-Ming Lee, and John Preskill}
\CALTECH
\ABSTRACT{We analyze the charges carried by loops of  string
in models with non-abelian local discrete symmetry.  The
charge on a loop has no localized source, but can be
detected by means of the Aharonov--Bohm interaction of the
loop with another string.  We describe the process of charge
detection, and the transfer of charge between point
particles and string loops, in terms of gauge--invariant
correlation functions.}

\ENDTITLEPAGE
\eject
\chapter{Introduction}
In a spontaneously broken gauge theory, if the unbroken
gauge group $H$ is a {\it discrete} subgroup of the
underlying continuous gauge group $G$, then the theory will
contain topologically stable strings (in 3+1 dimensions) or
vortices (in 2+1 dimensions).  If $H$ is non-abelian, the
strings have remarkable properties.  In particular, a closed
loop of string can carry a nontrivial $H$ charge.  Oddly,
this charge is a global property of the string that can not
be attributed to any locally defined charge density.  Yet
the charge is physically detectable, for the charged string
loop has an infinite range Aharonov--Bohm interaction with
other strings.  Furthermore, if a pointlike particle
carrying $H$ charge winds through a string loop, the
particle and the loop can exchange charge.

Charge with no localized source has been called ``Cheshire
charge.''~~\Ref\ABCMW{M. Alford, K. Benson, S. Coleman, J.
March-Russell, and F.
Wilczek, ``Interactions and Excitations of Non-Abelian
Vortices,''
Phys. Rev. Lett. {\bf 64}, 1632 (1990); {\bf 65,}
668 (1990) (E).}  It was first discussed for the case of the
``Alice'' string.\Ref\schwarz{A.S. Schwarz, ``Field Theories
With No Local Conservation of
Electric Charge,'' Nucl. Phys. {\bf B208}, 141 (1982);
A.S. Schwarz and Y.S. Tyupkin, ``Grand Unification and
Mirror Particles," Nucl. Phys. {\bf B209}, 427 (1982).}  A
loop of Alice string can carry electric  charge, and have a
long--range electric  field, even though the electric charge
density vanishes everywhere.\REF\ginsparg{P. Ginsparg and S.
Coleman, unpublished (1982).}\REF\prekra{J. Preskill and L.
Krauss, ``Local Discrete Symmetry and Quantum--Mechanical
Hair,'' Nucl. Phys. {\bf B341}, 50
(1990).}\refmark{\ginsparg,\ABCMW,\prekra}.
\REF\ABCMWmore{M. Alford, K. Benson, S. Coleman, J.
March-Russell, and F. Wilczek, ``Zero Modes of Non-Abelian
Vortices,'' Nucl. Phys. {\bf B349}, 414
(1991).}\REF\bulopr{M. Bucher, H.-K. Lo, and J. Preskill,
``Topological Approach to Alice Electrodynamics,'' Caltech
Preprint CALT-68-1752 (1991).}Processes in which electric
(or magnetic) charge is exchanged between string loops and
point particles were discussed in Ref.~\ABCMW,
\prekra-\bulopr.

In this paper, we analyze the purely quantum--mechanical
version of Cheshire charge that arises in a theory with a
non-abelian discrete local $H$ symmetry.\REF\AMW{M. Alford,
J. March-Russell, and F. Wilczek, ``Discrete Quantum Hair
on Black Holes and the Non-Abelian Aharonov--Bohm Effect,''
Nucl. Phys.
{\bf B337}, 695 (1990).}\refmark{\AMW,\prekra}  The
semiclassical theory of discrete Cheshire charge was
formulated in Ref.~\prekra, and elaborated in \REF\bucher{M.
Bucher, ``The Aharonov--Bohm Effect and Exotic Statistics for
Non--Abelian Vortices,'' Nucl. Phys. {\bf B350}, 163
(1991).}\REF\disentangling{M. Alford, S. Coleman and J.
March-Russell, ``Disentangling
Non--Abelian Discrete Quantum Hair,'' Nucl. Phys. {\bf
B351}, 735
(1991).}Ref.~\bucher, \disentangling.  Here we extend the
theory further, in several respects.  We describe how a
charge operator can be constructed, such that the
expectation value of the operator in a state specifies the
transformation properties of the state under global $H$
transformations.  We then study processes in which charge is
exchanged between string loops and point particles, and derive
general formulas for how the expectation value of the charge
of the loop is altered by the exchange.  Finally, we explain
how the charge exchange processes can be probed using
gauge--invariant correlation functions.

\REF\ALMP{M. Alford, K.-M. Lee, J. March-Russell, and J.
Preskill,
``Quantum Field Theory of Non-Abelian Strings and
Vortices,'' Caltech
preprint CALT-68-1700 (1991).}The charge operator and
correlation functions are also treated in Ref.~\ALMP, where
lattice realizations of operators and correlators are
extensively discussed.

The rest of this paper is organized as follows:  In Section
2, we briefly review the basic properties of non-abelian
strings and the concept of Cheshire charge.   We construct
the non-abelian charge operator in Section 3, and analyze
the charge exchange process in Section 4.  Section 5
contains a final comment.

\chapter{Non-Abelian Strings}
Let us briefly recall some of the properties of non-abelian
strings in three spatial dimensions (and vortices in two
spatial dimensions).

If a simply connected gauge group $G$ is broken to a
discrete subgroup $H$, then strings are classified by
elements of $H$.  To assign a group element to a loop of
string, we fix an (arbitrary) basepoint $x_0$, and specify a
path $C$, beginning and ending at $x_0$, that winds once
through the string loop.\FIG\path{The path $C$, starting and
ending at the point $x_0$, encircles a loop of string.}
(See Fig.~\path.)  The assigned group element is then
$$
a(C,x_0)=P~\exp\left(i\int_{C,x_0}A\cdot dx\right)~.
\eqn\stringA
$$
We refer to $a(C,x_0)$ as the ``flux'' of the string; it
encodes the effect of parallel transport around the path
$C$.  The flux takes values in $H(x_0)$, the subgroup of $G$
that stabilizes the Higgs condensate at the point $x_0$
(since parallel transport around $C$ must return the
condensate to its original value).  Since $H$ is discrete,
the flux $a(C,x_0)$ is unchanged by deformations of $C$ that
leave $x_0$ fixed, as long as $C$ never crosses the core of
the string.

\def\M{\cal M}
For a configuration of many string loops, we specify a
standard path for  each loop, where all paths have the same
basepoint.  Evidently, the flux associated with the product
path $C_2\circ C_1$ obtained by traversing first $C_1$ and
then $C_2$
is just the product $a(C_2,x_0)\cdot
a(C_1,x_0)$ of the two fluxes associated with $C_1$ and
$C_2$.  Thus, $a(C,x_0)$ defines a homomorphism that maps
$\pi_1({\M},x_0)$ to $H$, where $\M$ is the manifold that is
obtained when the cores of all strings are removed from
$\Re^3$.

The flux assigned to a path is not gauge invariant.
The gauge transformations at the basepoint $x_0$ that
preserve the condensate at the basepoint, and so preserve
the embedding of $H$ in $G$, take values in $H(x_0)$.  Under
such a gauge transformation  $h\in H(x_0)$, the flux
transforms as
$$
a(C,x_0)\to h~a(C,x_0)~h^{-1}~.
\eqn\stringB
$$
In a many--string configuration, the flux of each string
becomes conjugated by $h$.

In the presence of strings, the embedding of the unbroken
group $H$ in $G$ necessarily depends on the spatial position $x$.
If the strings are non-abelian, this position dependence is
described by a nontrivial fiber bundle.  The base space of
the bundle is the spatial manifold $\M$, the fiber is $H$,
and the structure group is also $H$, which acts on the fiber
by conjugation.  The bundle is twisted:  Upon transport
around the path $C$, the group element $h\in H(x_0)$ becomes
conjugated by $a(C,x_0)$.  This twist prevents the bundle
from being smoothly deformed to the trivial bundle $\M\times
H$.\REF\bala{A. Balachandran, F. Lizzi, and V. Rogers,
``Topological Symmetry Breakdown in Cholesterics,
Nematics, and ${}^3 He$,'' Phys. Rev. Lett. {\bf 52}, 1818
(1984).}  One thus says that the unbroken $H$ symmetry is
not ``globally realizable;''~\refmark{\bala,\ABCMW,\prekra}
there is no smooth function of position that describes how
the unbroken group is embedded in $G$.  Only the subgroup of
$H$ that commutes with the flux of all strings is globally
realizable on $\M$.

To define the $H$-charge of a state, we will want to
consider how the state transforms under global $H$
transformations.  Fortunately, these global gauge
transformations can be implemented, even though  a
topological obstruction prevents $H$ from being globally
realized.  The point is that it is sufficient to be able to
define an $H$ transformation on and outside a large surface
$\Sigma$ (homeomorphic to $S^2$)
that encloses all of the string loops.  The
transformation cannot be smoothly extended inside the sphere
if it is required to take values in $H(x)$.  However, one
may relax this requirement and allow
the gauge transformation to take values in
$G$ inside of $\Sigma$; then a smooth extension is possible.
It makes no difference what extension is chosen, for gauge
transformations of compact support act trivially on physical
states.  (In two spatial dimensions, the only global $H$
transformations that can be implemented are those that
commute with the {\it total}
flux; {\it i.e.}, the flux associated with a path that
encloses all of the vortices.)

If the basepoint $x_0$ lies outside the surface $\Sigma$,
then, under the global gauge transformation $h\in H$, the
flux of a string transforms as in eq.~\stringB.  Thus, the
$H$ representations mix up the string loop state labeled by
$a\in H$ with string loop states labeled by other group
elements in the same conjugacy class as $a$. Let
$[a]$ denote the conjugacy class that contains $a$.
The action of
$H$ on the members of the class $[a]$ defines a (reducible)
representation that we denote as $D^{([a])}$.  In
$D^{([a])}$, each element of $H$ is represented by a
permutation of the class, according to
$$
D^{([a])}(h):~\ket{a'}\to \ket{ha'h^{-1}}~,~~a'\in[a]~.
\eqn\stringC
$$
This representation can be decomposed into irreducible
representations of $H$.  For each class $[a]$ there is a
unique state that can be constructed that transforms
trivially under $H$; it is the superposition of flux
eigenstates
$$
\ket{0;[a]}={1\over\sqrt{n_{[a]}}}\sum_{a'\in[a]}\ket{
a'}~,
\eqn\stringD
$$
where $n_{[a]}$ denotes the order of the class.
The other states contained in the decomposition of
$D^{([a])}$ carry $H$-charge.  This is ``discrete
Cheshire charge,'' for the charge of the loop has no
localized source.  (Note that the charged string states transform trivially
under the {\it center} of $H$, since $D^{([a])}$ represents the
center trivially.)

The splitting between the charge-0 string state
eq.~\stringD\ and the lowest charge excitation of the string
is of order $\exp(-\kappa A)$, where $\kappa$ is a string
tension, and $A$ is the area of the string
loop.\refmark{\disentangling,\ALMP} It is a remarkable
property of Cheshire charge that, in the presence of a large
string loop, the gap between the ground state and the first
charged excitation is much less than the corresponding gap
when the string is absent. Indeed, the gap approaches zero
very rapidly as the size of the loop increases.

\chapter{Charge Operator}
The discrete charge of an object, including a charged string
loop, can be detected at long range by means of the
Aharonov--Bohm effect.\Ref\krawil{L. Krauss and F. Wilczek,
``Discrete Gauge Symmetry in
Continuum Theories,'' Phys. Rev. Lett. {\bf 62}, 1221
(1989).}
Let $\ket{u}$ denote the wave-function in
internal--symmetry space
of an object located at $x_0$ that transforms as the
irreducible representation $D^{(\nu)}$ of $H$.  Then when
the particle is transported around the closed path $C$ that
begins and ends at $x_0$, the wave-function is modified
according to
$$
\ket{u}\to D^{(\nu)}\left[a(C,x_0)\right]~\ket{u}~;
\eqn\stringE
$$
if the string is in the flux eigenstate $\ket{a}$, then the
Aharonov--Bohm phase that can be measured in an interference
experiment is
$$
\bra{u}D^{(\nu)}(a)\ket{u}~.
\eqn\stringF
$$
But if the string is in the charge--zero eigenstate
$\ket{0;[a]}$ given by eq.~\stringD, then the expectation
value of the ``phase'' $D^{(\nu)}(a)$ becomes
$$
{1\over n_{[a]}}\sum_{a'\in [a]} D^{(\nu)}(a')
={1\over n_H}\sum_{h\in H}D^{(\nu)}(hah^{-1})
={1\over n_\nu}\chi^{(\nu)}(a)~{\bf 1}~,
\eqn\stringG
$$
where $n_H$ is the order of the group, $n_\nu$ is the
dimension of $D^{(\nu)}$, and $\chi^{(\nu)}$ is the
character of
the representation. The second equality follows from Schur's
lemma.

In principle, the charge inside a large region can be
measured by means of a process in which the world sheet of a
string sweeps over the boundary of the region.  If the
string is in the charge-zero eigenstate $\ket{0;[a]}$,
and the object enclosed by the world sheet transforms as the
irreducible representation $(\nu)$ of $H$, then the
amplitude for this process will be weighted by the
Aharonov--Bohm factor $(1/n_\nu) \chi^{(\nu)}(a)$.  The
charge $(\nu)$ of an unidentified object can be determined
by measuring this factor for each class $[a]$.

A gauge--invariant operator $F_{[a]}(\Sigma)$ can be
constructed that inserts, as a classical source,  a string
worldsheet in the state $\ket{0;[a]}$ on the closed
surface $\Sigma$.  The realization of this operator in a
Euclidean lattice gauge theory was described in
Ref.~\prekra\ in the case where $H$ is abelian
\REF\fredenhagen{K. Fredenhagen, and M. Marcu, ``Charged
States in $Z_2$ Gauge Theories,'' Comm. Math.
Phys. {\bf 92}, 81 (1983); K. Fredenhagen, ``Particle
Structure of Gauge Theories,'' in {\it Fundamental Problems
of Gauge Field Theory}, ed. G. Velo and A. S. Wightman
(Plenum, New York, 1986).}\REF\AlMR{M. Alford and J.
March-Russell, ``New Order Parameters for
Non--Abelian Gauge Theories,'' Princeton preprint PUPT-1226
(1990).}(see also Ref.~\fredenhagen), and in
Ref.~\AlMR, \ALMP\ for $H$ non-abelian.  (It is closely
related to the 't Hooft loop operator.\Ref\Hooft{G. 't
Hooft, ``On the Phase Transition Towards Permanent
Quark Confinement,'' Nucl. Phys. {\bf B138}, 1 (1978);
``A Property of
Electric and Magnetic Flux in Non-Abelian Gauge Theories,''
Nucl.
Phys. {\bf B153}, 141 (1979).})  If the surface $\Sigma$ is
chosen to lie in a time slice, then the operator
$F_{[a]}(\Sigma)$ measures the charge enclosed by $\Sigma$.
To define the charge of an isolated object, we consider a
surface $\Sigma$ that encloses the object, and whose closest
approach to the object is large compared to the correlation
length of the theory.  Let $\ket{\psi}$ denote the quantum
state of the object.  Then we have
$$
{\bra{\psi}F_{[a]}(\Sigma)\ket{\psi}\over
\VEV{F_{[a]}(\Sigma)}_0}=
\sum_\nu p^{(\nu)}(\psi;\Sigma)~{1\over
n_\nu}\chi^{(\nu)}(a)~,
\eqn\stringH
$$
where $p^{(\nu)}(\psi;\Sigma)$ is the probability that the
object carries charge $(\nu)$.  By measuring
$F_{[a]}(\Sigma)$ for each class, we can determine all of
the $p^{(\nu)}$'s.  (It is necessary to divide by the vacuum
expectation value  $\VEV{F_{[a]}(\Sigma)}_0$ to remove the
effects of quantum--mechanical vacuum charge fluctuations
near the surface $\Sigma$.\refmark{\prekra})

The Aharonov-Bohm interaction makes it possible to detect
$H$-charge at arbitrarily long range.  Thus, a theory with
discrete local $H$ symmetry obeys a charge superselection
rule---no gauge--invariant local operator can create or
destroy $H$-charge.  We have
$$
\bra{(\mu)}{\cal O}\ket{(\nu)}=0~,~~~~(\mu)\ne (\nu)~,
\eqn\stringI
$$
where ${\cal O}$ is any local observable, and $\ket{(\nu)}$
denotes a state that transforms as the irreducible
representation $(\nu)$ of $H$.  We can construct a
projection operator that projects out a given superselection
sector of the Hilbert space.  It is
$$
P^{(\nu)}={n_\nu\over n_H}\sum_{a\in
H}\chi^{(\nu)}(a)^*~U(a)~,
\eqn\stringJ
$$
where $U(a)$ represents the global $H$ transformation $a\in
H$ acting on the Hilbert space.  This projection operator
can be expressed in terms of the operators $F_{[a]}(\Sigma)$, for it
follows from eq.~\stringH\ that
$$
{F_{[a]}(\Sigma)\over \VEV{F_{[a]}(\Sigma)}_0}~
\longrightarrow ~ {1\over n_{[a]}}\sum_{a'\in[a]}U(a')~,
\eqn\stringK
$$
as the surface $\Sigma$ approaches the surface at spatial
infinity.

    We can also use the operator $F_{[a]}$ to construct an
``Aharonov--Bohm Order Parameter'' (ABOP) that probes
whether nontrivial superselection sectors actually exist.
Let
$$
W^{(\nu)}(C)\equiv \chi^{(\nu)}\left[
P~\exp\left(i\int_C A\cdot dx\right)\right]
\eqn\stringL
$$
denote the Wilson loop operator in the irreducible
representation $(\nu)$.  This operator introduces a
classical source with charge $(\nu)$ propagating on the
world line $C$.  The ABOP is defined by
$$
A_{[a]}^{(\nu)}(\Sigma,C)\equiv
{F_{[a]}(\Sigma)~W^{(\nu)}(C)\over
\VEV{F_{[a]}(\Sigma)}_0
\VEV{W^{(\nu)}(C)}_0}~.
\eqn\stringM
$$
If $H$ quantum numbers can indeed be detected at infinite
range, then we expect that
$$
\VEV{A_{[a]}^{(\nu)}(\Sigma,C)}_0~
\longrightarrow~{1\over
n_\nu}\chi^{(\nu)}(a^{k(\Sigma,C)})~,
\eqn\stringN
$$
in the limit in which $\Sigma$ and $C$ increase to infinite
size, with the closest approach of $\Sigma$ to $C$ also
approaching infinity.  Here $k(\Sigma,C)$ is the
linking number of the
surface $\Sigma$ and the loop $C$.  (In the abelian case,
the ABOP was first described in Ref.~\fredenhagen, and was
further elaborated in Ref.~\prekra.  The non-abelian
generalization was introduced in Ref.~\AlMR, and its
properties were extensively discussed in Ref.~\ALMP.)

The operators $F_{[a]}(\Sigma)$ and
$A_{[a]}^{(\nu)}(\Sigma,C)$ can also be constructed in two
spatial dimensions.  Then $\Sigma$ becomes a closed curve
that can be interpreted as the world line of a vortex--antivortex pair.

\chapter{Charge Transfer}
We will now consider the non-abelian Aharonov--Bohm
interactions between string loops and point particles, and
demonstrate that exchange of $H$-charge can occur.

The total $H$-charge of a composite system consisting of a
string loop and a charged particle can be measured by
studying the Aharonov--Bohm interaction  of the composite
with other, much larger, string loops.  Obviously, then, the
total $H$-charge of the composite must be conserved; it
cannot change when the particle winds through the loop.  Charge
exchange is an inevitable consequence of charge
conservation.

To see this, it is convenient to imagine a composite of a
string loop and a particle--antiparticle pair, where,
initially, both the loop and the pair have zero charge
(transform trivially under $H$).  Suppose that the particle
transforms as the irreducible representation $D^{(\nu)}$ of
$H$;
the antiparticle transforms as the conjugate representation.
Let $\{e^{(\nu)}_i~|~~i=1,2\dots n_\nu\}$ denote an
orthonormal basis for the vector space on which $D^{(\nu)}$
acts.  Then the initial state of
the pair has the group--theoretic structure
$$
\ket{\psi^{(\nu)}_{\rm in}}={1\over
\sqrt{n_\nu}}~\ket{e^{(\nu)*}_i\otimes e^{(\nu)}_i}~
\eqn\transA
$$
(summed over $i$).  The initial state of the loop is the
state $\ket{0;[a]}$ defined in eq.~\stringD.

Suppose that the particle and antiparticle are
initially at the point $x_0$.  Then
the particle traverses a path $C$ that winds through
the string loop and returns to $x_0$.
After this traversal, the
state of the pair and the state of the loop are correlated.
The total charge is still zero, but in general the pair
and the loop
both have a nontrivial charge.  We can infer the
final charge on the loop by calculating the final charge
carried by the pair.  In fact, the final charge of the pair
is actually independent of the initial charge of the loop;
it depends only on the class $[a]$.  Thus, to calculate
the final charge of the pair, we may take the state of the
loop to be the flux eigenstate $\ket{a}$ (where the flux
is defined in terms of the path $C$ as in eq.~\stringA).
It does not
matter how the class representative $a$ is chosen.

Using eq.~\stringE, we find that the state of the pair after
the traversal is
$$
\ket{\psi^{(\nu)}_{\rm fin},a}={1\over \sqrt{n_\nu}}~
\ket{e^{(\nu)*}_i\otimes e^{(\nu)}_j}
D^{(\nu)}_{ji}(a)~.
\eqn\transB
$$
This state $\ket{\psi^{(\nu)}_{\rm fin},a}$ does not
transform as a definite irreducible representation of $H$,
but it can, of course, be decomposed into states of definite
$H$-charge.  The probability $p^{(\mu)}$ that the $H$-charge
is $(\mu)$ can be extracted by using the projection operator
$P^{(\mu)}$ defined by eq.~\stringJ.  We find
$$
\eqalign{
p&^{(\mu)}_{\rm pair}(\nu,[a])=\bra{\psi^{(\nu)}_{\rm
fin},a}
P^{(\mu)}\ket{\psi^{(\nu)}_{\rm fin},a}\cr
&={n_\mu\over n_\nu n_H}\sum_{b\in H}\chi^{(\mu)}(b^{-1})
D^{(\nu)*}_{nm}(a)
\left\langle{e^{(\nu)*}_m\otimes e^{(\nu)}_n}
\Big|{e^{(\nu)*}_k\otimes e^{(\nu)}_l}
\right\rangle
D^{(\nu)*}_{ki}(b)D^{(\nu)}_{lj}(b)
D^{(\nu)}_{ji}(a)\cr
&={n_\mu\over n_\nu n_H}\sum_{b\in H}\chi^{(\mu)}(b^{-1})
\chi^{(\nu)}(bab^{-1}a^{-1})~.\cr}
\eqn\transC
$$
As we anticipated, this result
is unchanged if $a$ is replaced by $a'\in[a]$.

If the total $H$-charge is zero, then the composite of
string loop and pair has a wavefunction of the form
$$
\ket{\psi^{(\nu)}_{[a]}}=\sum_\mu\sqrt{p^{(\mu)}_{\rm
pair}}~
\ket{{\rm loop},\mu *}\otimes\ket{{\rm pair},\mu}~.
\eqn\transD
$$
Thus, the probability that the loop carries charge $(\mu)$
is given by
$$
p^{(\mu)}_{\rm loop}(\nu,[a])=p^{(\mu*)}_{\rm pair}(\nu,[a])
=p^{(\mu *)}_{\rm loop}(\nu *,[a])
=p^{(\mu *)}_{\rm loop}(\nu,[a^{-1}])~.
\eqn\transE
$$
Of course, this probability is nonvanishing only if $D^{(\mu)}$
is contained in $D^{(\nu)*}\otimes D^{(\nu)}$ and represents the
center of $H$ trivially.

\FIG\borromean{The Borromean rings.  $C_1$ is the world line
of an $[a]$ vortex, $C_2$ is the world line of a $[b]$
vortex, and $C_3$ is the world line of a charged particle
that transforms as the representation $(\nu)$.  The charged
particle transfers charge to the $[a]$ vortex--antivortex
pair, and the charge is subsequently detected via the
Aharonov--Bohm interaction of the pair with the $[b]$
vortex.}
We can directly verify that detectable Cheshire charge now
resides on the string loop by studying an appropriate
gauge--invariant correlation function.  Consider the process
depicted in Fig.~\borromean.  This process is shown in 2+1
dimensions for ease of visualization, but the generalization
to 3+1 dimensions is straightforward.  At time $t_1$, a
vortex--antivortex pair is created.  The flux of the vortex
lies in the class $[a]$, and the (initial) $H$-charge of the
vortex pair is trivial.
At time $t_2$, a particle--antiparticle pair is created.
The particle has $H$-charge  $(\nu)$, and the pair is
(initially) uncharged.  Then the particle winds
counterclockwise
around the $[a]$ vortex, transferring charge to the vortex
pair.  Next, another vortex--antivortex pair, with flux
lying in the class $[b]$, winds around the (now charged)
$[a]$ vortex pair, acquiring an Aharonov--Bohm phase that is
sensitive to the charge of the $[a]$ pair.  Then the
charge-$(\nu)$ particle winds clockwise around the
$[a]$ vortex, discharging the $[a]$ pair.  Finally, the
particle--antiparticle pair is annihilated at time $t_3$,
and the $[a]$ vortex--antivortex pair is annihilated at time
$t_4$.

If the vortices and charged particles are treated as
classical sources, this process is described by the
correlation function
$$
\VEV{F_{[a]}(C_1)~F_{[b]}(C_2)~W^{(\nu)}(C_3)}_0~,
\eqn\transF
$$
where $C_1$ is the world line of the $[a]$ vortex, $C_2$ is
the world line of the $[b]$ vortex, and $C_3$ is the world
line of the charged particle.  As shown in Fig.~\borromean,
the three loops $C_1$, $C_2$, and $C_3$ are joined in a
topologically nontrivial configuration known as the
``Borromean rings;''~\Ref\links{See, for example, D.
Rolfsen, {\it Knots and Links} (Publish or Perish,
Wilmington, 1976).}  no two loops are linked, yet the loops
cannot be separated without crossing.

\FIG\loops{A deformation of the rings shown in
Fig.~\borromean.
The gauge field is singular on the
surfaces $S_1$ and $S_2$ that are bounded by
the loops $C_1$ and $C_2$.}
By considering the case where the loops are large and far
apart, and comparing with the case where the loops are
unjoined, we can isolate the Aharonov--Bohm factor acquired
by the $[b]$ vortex pair that winds around the charged $[a]$
vortex pair.  The calculation of eq.~\transF, using
weak--coupling perturbation theory on the lattice, is described in
Ref.~\ALMP.  We will not repeat the details of the
calculation here, but it is easy to explain the main idea.
Loosely speaking, inserting a classical vortex with flux $a$
on the closed path $C_1$ is equivalent to performing a
singular gauge transformation on a surface $S_1$ that is
bounded by $C_1$.  The path has an orientation, which induces
an orientation of the surface. The effect of the singular gauge
transformation on the Wilson loop $W^{(\nu)}(C_3)$ is to
insert the factor $D^{(\nu)}(a)$ where $C_3$ crosses $S_1$
in a positive sense, and to insert the factor $D^{(\nu)}(a^{-1})$
where $C_3$ crosses $S_1$ in a negative sense.
In Fig.~\loops, we see that the loop $C_3$ successively crosses
$S_2$ in a negative sense,  $S_1$ in a negative sense, $S_2$ in a
positive sense, and
$S_1$ in a positive sense, before closing.  Due to the path
ordering of the Wilson loop, the factor due to a later
crossing appears to the left of the factor due to an earlier
crossing,
These
crossings therefore modify $\VEV{W^{(\nu)}(C_3)}_0$ by the factor
$(1/n_\nu)\chi^{(\nu)}(aba^{-1}b^{-1})$ compared to the case
where $C_3$ is unjoined with $C_1$ and $C_2$.  Recalling
that $a$ and $b$ are averaged over a class when $F_{[a]}$
and $F_{[b]}$ are inserted, we find that
$$
{\VEV{F_{[a]}(C_1)~F_{[b]}(C_2)~W^{(\nu)}(C_3)}_0\over
\VEV{F_{[a]}(C_1)}_0\VEV{F_{[b]}(C_2)}_0\VEV{W^{(\nu)}(C_3)}_0}
{}~\longrightarrow~
{1\over n_H}\sum_{h\in H}
{1\over n_\nu}\chi^{(\nu)}(hah^{-1}bha^{-1}h^{-1}b^{-1})
\eqn\transG
$$
when the loops are large, far apart, and joined.

In 3+1 dimensions, there is an analog of the Borromean ring
configuration, in which two disjoint closed surfaces
$\Sigma_1$ and $\Sigma_2$ are joined by a closed loop $C_3$
that does not link with either surface.  For this
configuration, eq.~\transG\ still applies, with $C_1$ and
$C_2$ replaced by $\Sigma_1$ and $\Sigma_2$.
We can decompose the right-hand-side of eq.~\transG\ into
characters as
$$
\sum_\mu p^{(\mu)}_{\rm loop}(\nu,[a])~{1\over
n_\mu}\chi^{(\mu)}(b)~,
\eqn\transH
$$
where $p^{(\mu)}_{\rm loop}(\nu,[a])$ is the probability
that the charge carried by the $[a]$ string loop, and
detected by the $[b]$ string loop, is $(\mu)$.  (Compare
eq.~\stringH.)  Using the orthogonality of the characters,
we find from eq.~\transG\ and \transH\ that
$$
p^{(\mu)}_{\rm loop}(\nu,[a])={n_\mu\over n_\nu n_H}
\sum_{b\in H}\chi^{(\mu)}(b^{-1})\chi^{(\nu)}(aba^{-1}b^{-1})~,
\eqn\transI
$$
in agreement with eq.~\transE\ and \transC.  Thus, the
charge lost by the particle pair has indeed been transferred
to the $[a]$ string loop.   (Note that,
in order to get the right
answer, it is important to choose consistent orientations for the
world sheets $\Sigma_1$ and $\Sigma_2$---the $[a]$ string must pass
through the $[b]$ string in the same sense that the Wilson loop
passes through the $[a]$ string.  Otherwise, we would in effect
be measuring the charge of the $[a]$ string with a $[b^{-1}]$
string, rather than a $[b]$ string.)

We will now derive eq.~\transG\ by a different method that
invokes the ``holonomy interaction'' between string loops.
Consider two flux--eigenstate string loops that initially
carry flux $a$ and $b$.  Now suppose that the $b$ loop
sweeps around the $a$ loop and returns to its original
position.  After this process, the flux of the $b$ loop is
unchanged, but the flux of the $a$ loop has been altered; it
has become a $bab^{-1}$ loop.\REF\wu{F. Wilczek and Y.-S.
Wu, ``Space-Time Approach to Holonomy Scattering,''
Phys. Rev. Lett. {\bf 65}, 13
(1990).}\refmark{\wu,\bucher,\bulopr}  (Here again, we must
be careful about the orientations of the string loops.  The $a$
loop becomes a $bab^{-1}$ loop if it passes through the $b$ loop
in the same sense as the path $C$ that is used to define the flux
of the $b$ loop.  If it passes through the $b$ loop in the opposite
sense, it becomes a $b^{-1}ab$ loop.)

Return now to the Borromean ring process.  Suppose that two
string loops are initially in the flux eigenstate
$\ket{a,b}$.  Then a particle--antiparticle pair is created,
and the particle winds through the $a$ loop; the new state
of the string loops and the particle--antiparticle pair can
be expressed as
$$
{1\over \sqrt{n_\nu}}~
\ket{a,b,e^{(\nu)*}_i\otimes e^{(\nu)}_j}
D^{(\nu)}_{ji}(a)~.
\eqn\transJ
$$
(Compare eq.~\transB.)  When the $b$ loop sweeps around the
$a$ loop, the state becomes
$$
{1\over \sqrt{n_\nu}}~
\ket{bab^{-1},b,e^{(\nu)*}_i\otimes e^{(\nu)}_j}
D^{(\nu)}_{ji}(a)~,
\eqn\transK
$$
due to the holonomy interaction.
Now the particle winds back through the $bab^{-1}$ loop (in
the opposite sense), and the state becomes
$$
{1\over \sqrt{n_\nu}}~
\ket{bab^{-1},b,e^{(\nu)*}_i\otimes e^{(\nu)}_k}
D^{(\nu)}_{kj}(ba^{-1}b^{-1})
D^{(\nu)}_{ji}(a)~.
\eqn\transL
$$
Finally, the particle--antiparticle pair annihilates, and we
have
$$
{1\over n_\nu}\chi^{(\nu)}(aba^{-1}b^{-1})
\ket{b^{-1}ab,b}~.
\eqn\transM
$$

To reproduce eq.~\transG, we must
take the initial string state to be $\ket{0;[a],~0;[b]}$,
in which the $[a]$ and $[b]$ loops are both uncharged.
Thus, we average both $a$ and $b$ over a class.  We find
that the effect of the particle--antiparticle pair on the
string state is
$$
\ket{0;[a], ~0;[b]}\longrightarrow
\left({1\over n_H}\sum_{h\in H}
{1\over n_\nu}\chi^{(\nu)}(hah^{-1}bha^{-1}h^{-1}b^{-1})
\right)~\ket{0;[a],~0;[b]}~.
\eqn\transN
$$
By creating the initial string state and annihilating the
final string state, we obtain eq.~\transG.

\chapter{A Final Comment}
We described in Section 3 how a charge--zero string loop can
be used in an Aharonov--Bohm interference experiment to
measure the charge of an object.  (The corresponding
measurement process, using flux eigenstate strings, was
described in Ref.~\disentangling.)  We can imagine doing a
double--slit experiment with a beam of particles of unknown
charge, where a string loop in the state $\ket{0;[a]}$
surrounds one of the slits.  By observing how the shift in
the interference pattern depends on  the class $[a]$, we
can determine the character of the representation according
to which the particles in the beam transform, and so infer
their charge.

However, the phenomenon of charge transfer raises a puzzle.
If a particle passes through the slit that is surrounded by
the string, it transfers charge to the string.  By measuring
the charge on the string loop later, we can find out which
slit the particle passed through.  Thus, no interference
pattern should be seen.

The resolution of this puzzle is that there is a
nonvanishing probability, in general, that no charge
transfer takes place.  This probability is given by
eq.~\transI\ in the case where $(\mu)$ is the trivial
representation $(0)$; we then have
$$
p^{(0)}_{\rm loop}(\nu,[a])=\left|{1\over n_\nu}
\chi^{(\nu)}(a)\right|^2~.
\eqn\conclA
$$
Therefore, as long as the character does not vanish, it is
possible for the particle to slip through the string loop
without being detected, and an interference pattern is
observed.  From the interference pattern, the phase of the
character, as well as its modulus, can be deduced.

\bigskip
We thank Mark Alford, Hoi-Kwong Lo, John
March-Russell, and David Wales for useful discussions.
\refout
\figout
\bye